\documentstyle[12pt]{article}
\begin{document}
\title{
{\LARGE {\bf 
The gravitational and electroweak interactions unified as
a gauge theory of the de Sitter group}}}
\author{
{\normalsize {\bf Nikolaos A. Batakis\thanks
{electronic address: nbatakis@cc.uoi.gr}}}\\ 
{\normalsize Department of Physics, University of Ioannina} \\
{\normalsize GR-45110 Ioannina, Greece}}
\date{}
\maketitle
\begin{abstract}
\begin{sloppypar}
\normalsize

\noindent
The complexified gauging of the de Sitter group gives
a unified theory for the electroweak and gravitational 
interactions. The standard
spectrum for the electroweak gauge bosons is recovered with
the correct mass assignments, following a
spontaneous breaking of the gauge symmetry 
imposed by the geometry. There is no conventional
Higgs sector. New physics is predicted with gravity-induced 
electroweak processes (at the electroweak and 
at an intermediate scale of about $10^{10}Gev$) 
as well as with novel-type of effects
(such as gravitational Aharonov-Bohm and violations of the 
Principle of equivalence to 1 part in $10^{17}$). The new 
theoretical perspectives emerging from this geometric
unification are briefly discussed.

\vspace{1cm}
\noindent
PACS numbers: 03.50z, 04.50.+h, 04.80.Cc, 04.90.+e,
11.15.-q, 11.15.Ex, 11.30.-j, 11.30.Qc, 12.10.-g,
12.60.-i, 12.60.Cn, 12.90.+b. 

\end{sloppypar}
\end{abstract}
\addtolength{\baselineskip}{.3\baselineskip}
\newpage 

\section{Introduction}

Following the early \cite{1} up to the contemporary 
\cite{2g} efforts to unify 
gravity with other fundamental interactions,
one is tempted to isolate as perhaps the single most
important finding the fact that Einstein's 4D 
theory can also be formulated as a gauge theory 
of the Poincar\'{e} group ${\cal P}_0$ \cite{4}. We recall
that under the conventional (that is to 
say, {\em real}) gauging of ${\cal P}_0$,
Einstein's gravity emerges as expected, 
namely associated with the subgroup ${\cal L}$
of Lorentz rotations. Curiously, however, 
there is no gauge-field output
corresponding to the translational generators, for which
we only seem to have a dubious association with spin \cite{rc}.
Updating earlier attempts and refining recent preliminary results
\cite{nb}, we will expand here on the premise that
the translational generators of space-time are intimately
associated with the electroweak interaction.
We will see that the {\em complexified} gauging of ${\cal P}$,
a uniquely defined 10-parameter Lie group which turns out to be
isomorphic to the de Sitter group, gives rise to 
a spontaneously broken $SU(2)\times U(1)$ gauge theory
unified with gravity.
In the following section we will introduce ${\cal P}$
and examine its conventional gauge theory.
In section 3 we will proceed with 
the {\em complexified} gauging of ${\cal P}$
in the context of Riemann-Cartan geometry, but without
commitment on any particular
spin-torsion interrelation. In
section 4 we will trace the intricate passage
from the space-time geometry to
an internal gauge symmetry. 
In section 5 we will
uncover and examine an already present mechanism
for the spontaneous breaking of that symmetry:
it is elegantly 
imposed by the geometry without any {\em ad hoc} assumptions,
so the need of a Higgs sector is thus superseded.
In section 6 we will examine the relevance of our findings
to the electroweak interaction \cite{3}, to realize that 
we actually recover
the standard mass spectrum and charges
for the electroweak gauge
bosons. Our construction seems to 
offer fundamental upgrading
to each one of the two interactions it unifies, together with new
(and apparently testable) predictions, 
as we will discuss in the last section.

To establish notation \cite{4}, we briefly recall that a
contemporary formulation of Einstein's theory in a
differentiable 4D manifold $M^4$ 
involves the Einstein-Hilbert action as
\begin{equation}
I_{EH} =\frac{1}{32\pi G}
\int_{M^4}R_{ab}\wedge\ast(e^a\wedge e^b),
\label{eh}
\end{equation}
wherefrom the vacuum field equations follow upon 
variation of the frame $e^a$ under the requirement
of $SO(1,3)$ 
(recast from the original $GL_4$ co-ordinate) invariance,
and the constraints of vanishing torsion and metricity.
The latter is by definition expressed as 
$Dg_{ab}=0$ which, with $g_{ab}=(-1,1,1,1)$,
amounts to antisymmetry
of the connection $\omega^{ab}$.

We also recall the alternative formulation of Einstein's gravity
as a gauge theory of the Poincar\'{e} group 
with generators $Q_A \equiv\{P_a,M_{aa'}\}$ 
assigned, respectively, to
the translations and the ${\cal L}$ rotations in $M^4$. According
to the standard procedure, we gauge ${\cal P}_0$ (and, eventually,
${\cal P}$) with the introduction of the generalized potential
\begin{equation}
H\equiv H^AQ_A=e^aP_a+\frac{1}{2}\omega^{ab}M_{ab},
\label{H}
\end{equation}
with covariant derivative $D$ (cf. below), and field strength
\begin{equation}
\Omega\equiv\Omega^AQ_A=dH+H\wedge H=
(dH^A+\frac{1}{2}f^A_{BC}H^B\wedge H^C)Q_A=
T^aP_a+\frac{1}{2}R^{ab}M_{ab}.
\label{Om}
\end{equation}
To explicitly write down the torsion $T^a$ and
curvature $R^{ab}$ 2-forms 
(namely Cartan's structure equations), we need the
structure constants $f^A_{BC}$. 
Utilizing the algebra of ${\cal P}_0$, we find 
\begin{equation}
\stackrel{\circ}{T}{^a}=
\stackrel{\circ}{D}
e^a\equiv de^a+
\stackrel{\circ}
{\omega}{_{\,\cdot b}^{\,a}}\wedge e^b,\;\;\;\;\;
\stackrel{\circ}{R}{^{ab}}=
d\stackrel{\circ}
{\omega}{^{ab}}+
\stackrel{\circ}
{\omega}{^a_{\cdot c}}\wedge
\stackrel{\circ}
{\omega}{^{cb}},
\label{c0}
\end{equation}
for the gauge theory of the Poincar\'{e} group.
Then, variation of the Einstein-Hilbert action 
(in the absence of any external
spinorial sources) with respect to $\omega^{ab}$ 
(24 independent components) sets to zero 
the 24 independent components of
the torsion, while variation with 
respect to $e^a$ furnishes Einstein's
equations in vacuum. 

We finally recall the de Sitter group \cite{ds}, with its 
algebra expressed as
\begin{equation}
[P_a,P_b]=-\beta M_{ab},\;\;
[M_{aa'},P_b]=2\delta^c_{[a}g_{a']b}P_c,\;\;
[M_{aa'},M_{bb'}]=4\delta^c_{[a} g_{a'][b}\delta^{c'}_{b']}M_{cc'},
\label{DS} 
\end{equation}
where the $\pm |\beta|$ values of the (real) parameter $\beta$
are associated with the two possible types of
non-trivial topology involved (closed and open for positive
and negative curvature),
while the $\beta=0$ value practically yields
the standard contraction down to the 
Poincar\'{e} algebra.
Cartan's structure equations for the above basis of
the de Sitter agebra are  
\begin{equation}
\stackrel{1}{T}{^a}=
\stackrel{1}{D}
e^a\equiv de^a+
\stackrel{1}
{\omega}{_{\,\cdot b}^{\,a}}\wedge e^b,\;\;\;\;\;
\stackrel{1}{R}{^{ab}}=
d\stackrel{1}
{\omega}{^{ab}}+
\stackrel{1}
{\omega}{^a_{\cdot c}}\wedge
\stackrel{1}
{\omega}{^{cb}}-
\beta e^a\wedge e^b.
\label{cds}
\end{equation}
Comparing (\ref{c0},\ref{cds}) we conclude that, 
besides the {\em global} distinction
(from topology),
the only difference 
(at the classical level)
between the Poincar\'{e} and the de Sitter
gauge theory is
a comological constant proportional
to $\beta$. In particular, there is again no 
field-strength output corresponding to the $P_a$
generators: in vacuum, the torsion has to vanish
identically.

\section{Introducing ${\cal P}$: uniqueness,
isomorphism to the de Sitter group, and gauging}

The mentioned peculiarity of having no field strength
associated with the translational generators can actually
be traced to much deeper issues such as the nature of
torsion and its relation to spin and the quantization
of gravity. The apparent impass is enhanced by the
Coleman-Mandula theorem \cite{cm} plus the fact that the
de Sitter (together with its contraction
to the Poincar\'{e}) group exhausts all three
possibilities for the isometry groups of maximally
symmetric 4D manifolds. The best known resolutions
proposed include the twistor approach
and, of course, supersymmetry \cite{2g}. 
Reflecting on the above, we have been motivated to look
for a generic 10-parameter 
Lie group ${\cal P}$ (with generators $\Pi_a,M_{aa'}$), 
not necessarily constrained to be a maximal-isometry
group, but which must contain ${\cal L}$ 
as a subgroup. This would
imply that, although ${\cal P}$
will certainly act {\em transitively}
on ${\cal P}/{\cal L}$, it might {\em not} do so
on space-time and, in particular, its $\Pi_a$ generators
may loose a direct interpretation as space-time
translations. The latter will, of course, exist and be
well defined in any case,
along with whatever symmetry they may have.

Let us then formally introduce ${\cal P}$, with its generators
satisfying the algebra
\begin{eqnarray}
&&[\Pi_a,\Pi_b]=C^c_{ab}\Pi_c+
C^{cc'}_{ab}M_{cc'}, \nonumber \\
&&[M_{aa'},\Pi_b]=2\delta^c_{[a}g_{a']b}\Pi_c+
C^{cc'}_{aa'\,b}
M_{cc'}, \nonumber \\
&&[M_{aa'},M_{bb'}]=4\delta^c_{[a}
g_{a'][b}\delta^{c'}_{b']}M_{cc'}.
\label{PP} 
\end{eqnarray}
The acceptable
choices for such a {\em simple} algebra 
are severely restricted. In fact, the 
Jacobi identity dictates
that $C^a_{bc}$ are precisely the
structure constants for the algebra
of $SU(2)\times U(1)$, namely they are numerically
equal to $\epsilon_{jkl}$ for $j,...=\{1,2,3\}$ or 
zero if any one of the $a,b,c$ indices is zero, while
for the remaining structure constants we find
\begin{equation}
C^{cc'}_{ab}=\frac{1}{8}\left(C^d_{ab}C^{cc'}_{\;\cdot d}
-\delta_a^c\delta_b^{c'} 
+\delta_b^c\delta_a^{c'}\right), \;\;\;\;\;
C^{cc'}_{aa'\,b}=\delta^{[c}_{[a}C_{a']b}^{c']}+
\frac{1}{2}C_{\;\cdot[a}^{cc'}g_{a']b}.
\label{PPC} 
\end{equation}
It follows that ${\cal P}$ contains, in addition to ${\cal L}$,
a distinct $SU(2)$ subgroup generated by
$\Pi_j$ and a $U(1)$ associated with $\Pi_0$.
To explicitly see this structure, we introduce the
usual $J_j=-\frac{1}{2}\epsilon_{jkl}M_{kl}$,
$K_j=M_{0j}$ generators (spatial rotations and Lorentz boosts), 
so that the commutation relations (\ref{PP}) may equivalently
be written as
\begin{eqnarray}
&&[J_j,J_k]=\epsilon_{jkl}J_l,\;\;
[J_j,K_k]=\epsilon_{jkl}K_l,\;\;
[J_j,\Pi_k]=\epsilon_{jkl}\Pi_l,\;\;
[J_j,\Pi_0]=0,
\nonumber \\
&&[K_j,K_k]=-\epsilon_{jkl}J_l,\;\;
[K_j,\Pi_k]=g_{jk}\Pi_0+\frac{1}{2}\epsilon_{jkl}K_l,\;\;
[K_j,\Pi_0]=\Pi_j-\frac{1}{2}J_j,
\nonumber \\
&&[\Pi_j,\Pi_k]=\epsilon_{jkl}\Pi_l,\;\;
[\Pi_j,\Pi_0]=\frac{1}{4}K_j.
\label{ccp}
\end{eqnarray}
In this basis, the $SU(2)$ subgroup structure 
generated by the $\Pi_j$ is
obvious, and we also note that the commutator of each
$\Pi_j$ with $\Pi_0$ does not vanish but closes
to a Lorenz boost along $j$.

Further investigation has shown that 
${\cal P}$ is, in fact, unique,
{\em modulo} isomorphisms and trivial cases such as
direct products of 4-parameter Lie groups
with ${\cal L}$ or contractions of ${\cal P}$ 
(e.g., down to the Poincar\'{e} group).
The de Sitter group, in
particular, is actually isomorphic
to ${\cal P}$. This can be established
if one introduces the new set of (translational)
generators $P_a$ with 
\begin{equation}
P_a=2\sqrt {\beta}\left(\Pi_a+
\frac{1}{4}C^{cc'}_{\;\cdot a}M_{cc'}\right)\;\;
\;\;\leftrightarrow\;\;
\;\;P_0=2\sqrt {\beta}\Pi_0,\;
P_j=2\sqrt {\beta}\left(\Pi_j-\frac{1}{2}J_j\right),
\label{dsr}
\end{equation}
to finally show that the 
$P_a,M_{aa'}$ set 
satisfies precisely the de Sitter algebra (\ref{DS}).

We may now proceed with the (real) gauging of ${\cal P}$,
utilizing the basis in (\ref{PP}) with 
components of the generalized potential
$(e^a,\frac{1}{2}\stackrel{2}{\omega}{^{aa'}})$, to find
Cartan's equations as
\begin{eqnarray}
\stackrel{2}{T}{^a}=T^la &=& de^a+
\frac{1}{2}C^a_{bc}e^b\wedge e^c
+\stackrel{2}
{\omega}{_{\,\cdot b}^{\,a}}\wedge e^b,
\label{c} \\
\stackrel{2}{R}{^{ab}}=R^{ab} &=&
d\stackrel{2}
{\omega}{^{ab}}+
\stackrel{2}
{\omega}{^a_{\cdot c}}\wedge
\stackrel{2}{\omega}{^{cb}}+
\frac{1}{2}\left(C^a_{cd} 
\stackrel{2}{\omega}{^{cb}}+
C^b_{cd}
\stackrel{2}{\omega}{^{ac}}+
C^{ab}_{\;\cdot c}
\stackrel{2}
{\omega}{^c_{\cdot d}}\right)\wedge e^d+
C^{ab}_{cd}e^c\wedge e^d,
\nonumber
\end{eqnarray}
with $C^{ab}_{cd}$ given by (\ref{PPC}).
Due to the established isomorphism between
${\cal P}$ and the de Sitter group, 
one might expect that the content of the
sets (\ref{cds}) and (\ref{c}) is identical.
In fact, this is not true because
of the different gauging involved in each case.
To better compare the two sets, we may re-cast
(\ref{c}) as
\begin{equation}
\stackrel{2}{T}{^a}=
De^a\equiv de^a+
\omega^a_{\cdot b}\wedge e^b,\;\;\;
\;\;\stackrel{2}{R}{^{ab}}=
d\omega^{ab}+
\omega^a_{\cdot c}\wedge\omega^{cb}
-\frac{1}{4}e^a\wedge e^b
\frac{1}{2}C^{ab}_{\;\cdot c}De^c,
\label{b} 
\end{equation}
expressed in terms of the new connection
\begin{equation}
\omega^{ab} \equiv
\stackrel{2}{\omega}{^{ab}}
-\frac{1}{2}C^{ab}_{\;\cdot c}e^c,
\label{bc} 
\end{equation}
and its respective covariant derivative $D$.
As compared to the
gauge theory of the de Sitter group (namely
(\ref{cds}) etc.), the above equations
supply a formally identical expression for the torsion,
but there are differences in the curvature. Besides the
modification in the contribution
for the cosmological constant, 
the presence of the additional term proportional to 
$De^a$ (namely to
the torsion) is non-redundant and cannot be
transfomed or gauged away, The reason is that, unlike the 
case for the de Sitter group,
variation of the Einstein-Hilbert action
(written in terms of either
of the expressions (\ref{c},\ref{b}))
with respect to $\stackrel{2}{\omega}{^{aa'}}$ shows that
the torsion cannot vanish
in the gauge theory of ${\cal P}$, in
fact it turns out to be
\begin{equation}
{T}{^a}=
\frac{1}{6}C^a_{bc}e^b\wedge e^c.
\label{bt} 
\end{equation}
We observe that we now have a non-vanishing
field strength associated with the $\Pi_a$
generators. To understand the diference in view of
the identical expressions for the torsion
in (\ref{cds}, \ref{b}), we observe that
these expressions are definitions of $T^a$, while 
(\ref{bt}) is its field equation (whose counterpart
in the de Sitter case is $T^a=0$).
Obviously, this field equation is an algebraic one, in accord
with the general result that
torsion does not propagate \cite{4}.
In fact, we will eventually also get a propagating
field (the electroweak)
associated with the $\Pi_a$,
following the
complexified gauging of ${\cal P}$.

\section{Complexified gauging of ${\cal P}$}
 
By complexified gauging of ${\cal P}$ we simply
mean that we now let $H$ 
in (\ref{H}-\ref{Om})
become complex, namely we re-write (\ref{c})
or (\ref{b}) with the substitutions 
\begin{equation}
e^a\rightarrow\tilde{e}^a=e^a+ih^a,\;\;
\;\;\;\omega^{ab}\rightarrow
\tilde{\omega}^{ab}=
\omega^{ab}+
i\left(K^{ab}+S^{ab}\right),
\label{et}
\end{equation}
or, equivalently,
\begin{equation}
e^a\rightarrow\tilde{e}^a=e^a+ih^a,\;\;
\;\;\;\stackrel{2}{\omega}{^{ab}}\rightarrow
\stackrel{2}{\omega}{^{ab}}+
i\left(\frac{1}{2}C^{ab}_{\;\cdot c}h^c+K^{ab}+S^{ab}\right),
\label{et2}
\end{equation}
with $\omega^{ab}$ still as defined by (\ref{bc}).
Before we proceed to discuss the physical
interpretation of this complexification, we note that
what is indicated as the imaginary part of the connection
in (\ref{et}) has been split into two pieces for later
convenience. In $K^{ab}=-K^{ba}$
we have chosen to segregate all
contributions coming from (and thus been determined by)
$h^a$, while $S^{ab}=-S^{ba}$ carries the needed 24
of the total 48 (real+imaginary) independent components
of $\tilde{\omega}^{ab}$. The antisymmetry has been
imposed so that metricity is maintained. 

At least formally, we could exploit the 
use of the complex covariant derivative
defined with  
$\tilde{\omega}^{ab}$ as a
complex connection.
The latter notion is already 
familiar from the treatment of non-abelian
gauge theory and connections are not observables
anyway, so there is really no problem
involved here.
On the other hand, this is not 
quite the case with 
$\tilde{e}^a$ (frames are geometrical objects),
so we will briefly examine the physical
interpretation associated with the
present employment of a complex frame. 

Let us recall
our earlier re-definition of a connection 
in (\ref{bc}), which fascilitated our
comparative examination of two gauge
theories, whereby a contribution
from the $e^a$ frame (the $\frac{1}{2}C^a_{bc}e^c$ term)
was effectively absorbed in the connection.
In the present context, the contribution of 
what is indicated in (\ref{et}) as the imaginary
part of $\tilde{e}^a$, namely $h^a$, 
can be likewise
absorbed in the already existing 
complex connection. As a result,
in addition to its gauging aspect, the above
complexification also admits the folowing
geometrical interpretation. Let us start
with the conventional (general relativistic)
description of $M^4$
given in terms of the {\em real}
basis $e^a$ and the also real
$\stackrel{2}{\omega}{^{ab}}$ as its 
{\em Christoffel} connection.
Following the complexification (\ref{et}), 
the quantities $e^a$ and 
$\stackrel{2}{\omega}{^{ab}}$ 
will retain both, their
reality as well as their previous identification
(even though the may change in value as a result of
additional sources in whatever will turn up
as Einstein's equations). However, the connection
of the resulting space-time will no longer be simply
Christoffel (just like the 
$\omega^{ab}$ in (\ref{bc}) is {\em not}).
In other words, 
$\stackrel{2}{\omega}{^{ab}}$ 
will be enlarged 
with a tensorial contribution,
namely with what is defined as {\em contorsion} in
the context of Riemann-Cartan geometry
(given by $-\frac{1}{2}C^a_{bc}e^c$ in (\ref{bc})). 
The only difference from that
context is that in our case the contorsion will be
complex: it will receive mixed (real+imaginary)
contributions from
$h^a$, $K^{ab}$, $S^{ab}$. The real part of this contorsion
will be observable through its conventional geometric
interpretation. The imaginary part will also
be observable, but in a gauge-theoretic 
(notably for the electroweak interaction)
and topological
(e.g., gravitational 
Aharonov-Bhom effects)
context. 

In the above analysis it
makes no difference if some or all of 
the $h^a$, $K^{ab}$, $S^{ab}$
become complex (e.g., as a result of the
employment of a particular representation or gauge).
Although no degrees of freedom will be gained or lost,
it will be useful to have
a clear perspective on that as well as
on potentials and their field strengths.
By the generic definitions 
in (\ref{H}-\ref{Om})
applied for the algebra (\ref{PP}), let
$\tilde{H}$ be
the new (complex) generalized potential
with field
strength $\tilde{\Omega}$.
The components of 
$\tilde{H}$ are specified as
\begin{equation}
\tilde{H}^a=e^a+ih^a,\;\;\;
\;\;\;\tilde{H}^{aa'}=
\frac{1}{2}\left(\stackrel{2}{\omega}{^{ab}}+
i\left(\frac{1}{2}C^{ab}_{\;\cdot c}h^c+
K^{ab}+S^{ab}\right)\right),
\label{H2}
\end{equation}
while those of $\tilde{\Omega}$, 
namely $\tilde{T}^a$, $\frac{1}{2}\tilde R^{ab}$),
will be calculated shortly.

The independent variables, namely those to be varied
in the classical action, are 16+24 for the real $e^a$
and $\stackrel{2}{\omega}^{aa'}$ 
plus an additional real count 16+24
coming from $h^a$ (actually $A^I$ in
terms of which $h^a$ will be defined - cf. below), 
and $S^{aa'}$. Due to certain calculational
subtleties, increased care is required in the choice of
the connection and its variation, otherwise 
one may end up with virtually unmanageable complexity.
In particular we note that our choice of the basis
in (\ref{PP}) and the connection (\ref{bc}), essentially 
a choice of gauge, may not be optimal.
Practically, the variation of 
$\omega^{aa'}$ seems preferable and equivalent to a  
variation of $\stackrel{2}{\omega}{^{aa'}}$.
However, in the former case, under the independent
variation $\delta e^a$ of the frame, one should
simultaneously vary the connection as
$\delta\omega^{ab}=-\frac{1}{2}C^{ab}_{\;\cdot c}
\delta e^c$. The rest of the idependent variables,
namely $S^{aa'}$, are expected to be associated
with the fermionic content of space-time but they will not
be really considered any further in the present treatment.

Re-tracing the steps leading to (\ref{c}) or (\ref{b})
(practically, just 
substituting (\ref{et}) in (\ref{b}), having
set $S^{ab}=0$) we find
Cartan's equations as
\begin{equation}
\tilde{T}^a=
d\tilde {e}^a+
\tilde{\omega}^a_{\cdot b}\wedge\tilde {e}^b,\;\;\;
\;\;\tilde{R}^{ab}=
d\tilde{\omega}^{ab}+
\tilde{\omega}^a_{\cdot c}\wedge\tilde{\omega}^{cb}
-\frac{1}{4}\tilde{e}^a\wedge \tilde{e}^b+
\frac{1}{2}C^{ab}_{\;\cdot c}\tilde{T}^c,
\label{tib} 
\end{equation}
These quantities specify directly observable field
strengths and the Lagrangian, so their imaginary parts
should vanish identically or behave as needed. In
particular they should not give
unacceptable imaginary contributions,
starting with the classical action. They
essentially carry the entire content
of the present theory (except for its symmetry breaking
aspect, introduced in section 5). Although one
can recognize the desired sectors (notably the
gravitational) already forming in (\ref{tib}), a rather
delicate handling is required for uncovering what actually 
is an internal gauge symmetry from space-time: the endeavoured
electroweak interaction from
the $SU(2)$ and $U(1)$ subgroups (generated by
$\Pi_j$ and $\Pi_0$, as we have seen).

\section{Internal symmetry extracted from space-time}
 
To proceed, we need explicit expressions 
for $h^a$ and $K^{ab}$ in (\ref{et}).
To uncover them, let us recall that 
the action of ${\cal P}$
on space-time may not be isometric, as mentioned,
but  $M^4$ may well be chosen to be
diffeomorphic  
to ${\cal P}/{\cal L}$. We will make this assumption
so that there exist in $M^4$ realizations
of the algebra of ${\cal P}$. 
This algebra contains, as we have seen, the two
sub-algebras for the $SU(2)$ and $U(1)$ subgroups
generated by ${\Pi_a}$. Although it is customary (in GUTs etc.)
to denote the presence of these subgroups 
as a direct product, we actually
have a semi-direct one in view of the 
$[\Pi_j,\Pi_0]$ commutator as given by (\ref{ccp}).
However, we have alredy seen
from (\ref{PP}) that the
structure constants $C^c_{ab}$ are presisely those of the
direct-product $SU(2)\times U(1)$ algebra, which
will be refered to as $u_2$. Next, we will
introduce a vierbein $\theta_J^a$ which will
relate vectors in the two 4D vector spaces, namely
$u_2$ and the tangent space at each point in $M^4$.

Let us then introduce in $M^4$ a set (frame) of four real 
vector fields 
$\theta_J=\theta_J^ae_a$,
ennumerated by the index $J=\{0,j\}=\{0,1,2,3\}$ and with
$e_a$ the dual of $e^a$.
The $\theta_J$ are chosen so that they satisfy the
commutation relations
\begin{equation}
[\theta_J,\theta_K]=C^I_{JK}\theta_I,\;\;\;\;\;
C^I_{JK}\equiv\theta^I_a\theta_J^b\theta_K^cC^a_{bc},
\label{com}
\end{equation}
where the commutators are defined as usual 
by the respective Lie derivatives 
and the $C^I_{JK}$ are defined in terms of
the $C^a_{bc}$ as indicated, with 
$\theta^I_a$ the matrix inverse of $\theta_I^a$.

We may now express
$h^a$ in (\ref{et}) in terms
of some $u_2$-valued 1-form $A^J=A^J_ae^a$ (where $A^J_a$
are differentiable functions in $M^4$) dotted to
the $\theta_J^a$. To do
that, we obviously
need an inner product between vectors in 
$u_2$. We are thus prompted to the definition
\begin{equation}
h^a\equiv\gamma^I_J\theta_I^aA^J=
(\cos \vartheta_W\theta_0^aA^0_b+
\sin \vartheta_W\theta_j^aA^j_b)e^b,
\label{dot}
\end{equation}
where the particular value chosen
for the constant matrix $\gamma^I_J$ has been generically
expressed (in view of the nature of the $u_2$ algebra) 
as a {\em mixing} by a single constant angle
paramerter $\vartheta_W$.

To re-express (\ref{tib}), we substitute
for $h^a$ its definition in terms of $A^I$,
which obviously does not change the real count of 24
independent variables. The result of this calculation is
\begin{equation}
\tilde{T}^a=T^a+\gamma^I_J\theta_I^bA^J\wedge K^a_b
+i\left(\gamma^I_J(D\theta^a_I)\wedge A^J+
\gamma^I_J\theta^a_I F^J +
K^a_b\wedge e^b\right), 
\label{tor}
\end{equation}
\begin{equation}
\tilde{R}^{ab}=R^{ab}
-K^a_c\wedge K^{cb}+
\frac{1}{2}C^{ab}_{\;\cdot c}\tilde {T}^c
+\frac{1}{4}\gamma^I_J \gamma^K_L
\theta^a_I \theta^b_K A^J\wedge A^L
+i\left(DK^{ab}-
\frac{1}{4}\gamma^I_J(\theta^a_Ie^b+
\theta^b_Ie^a)\wedge A^J\right). 
\label{R}
\end{equation}
In the above
expressions, we have made use of the definitions
\begin{equation}
D\theta_I^a \equiv  
d\theta_I^a+
\omega^a_b\theta_I^b
+\omega^J_I\theta_J^a,
\label{td} 
\end{equation}
\begin{equation}
F^I \equiv  dA^I
+\frac{1}{2}C^I_{JK}A^J\wedge A^K,
\;\;(\mbox{with} \;\theta_a^I\theta^b_I=\delta_a^b).
\label{f} 
\end{equation}
It should be noted that the definitions of
$T^a$ and $R^{ab}$ as given by (\ref{b}) are
clearly retained, although these quantitities will
acquire different values through the field
equations. The result (\ref{bt}), in particular, is
not expected to hold in the present context,
because it will be replaced by the new field equation
resulting from the variation of $\omega^{ab}$.
We also observe that the 
field strength $F^I$ is defined as usual for the
gauge potential $A^I$, while the
covariant derivative
$D$ remains the same, as long as
it does not meet a $u_2$ index, 
otherwise (as in (\ref{td})), there
is an extra term from
the A-connection in $u_2$, implemented by
\begin{equation}
\omega^I_J \equiv \frac{1}{2}C^I_{JK}A^K.
\label{acon} 
\end{equation}

For reasons which will be discussed shortly, we
will assume that the 
$\theta_I^a$ is covariantly constant, namely we will set 
$D\theta_I^a=0$ in (\ref{td}) (this may be viewed
as analogous to the $De^a=0$ in
conventional general relativity,
imposed there as the zero-torsion constraint.
Then, the indicated as imaginary
part of the torsion in (\ref{tor}) vanishes identically if
\begin{equation}
K^a_b\wedge e^b=
-\theta^a\cdot F.
\label{k}
\end{equation}
Reflecting upon the structure of (\ref{R}), we see
that a Yang-Mills sector associated with $F$ can
be expected to emerge automatically as
part of the Einstein-Hilbert action written for
$\tilde{R}^{ab}$. In realizing that, we will see
the space-time metric $g_{ab}$ transformed by the
vierbein $\theta^a_I$ into $g_{IJ}$ as
\begin{equation}
\theta_I^a \theta_J^b g_{ab}\equiv \frac{32\pi G}{g^2}g_{IJ},
\label{G}
\end{equation}
up to an overall factor specified by the parameter $g$.
The latter will, in fact, turn out to be the gauge 
coupling, which will obviously 
deviate from 
a constant to the extend that 
$D\theta^a_J=0$ is violated.
Such deviations would, in any case,
be negligible at energies
below those associated with the epoch of homogenization of the
universe. Moreover, to secure correct relative signs
in the Yang-Mills sector, 
the signature of
$g_{IJ}$ must be $\delta_{IJ}$ (all pluses),
attainable with a `Wick rotation' of $\theta_0$
to $i\theta_0$, hereafter re-defined as $\theta_0$,
by which all our results remain
formally intact.

We are now ready to write down the Einstein-Hilbert action
in $M^4$ with the
curvature tensor (\ref{R}).
Dropping the imaginary surface term and the gauge-fixing terms
(which, however, would be of importance
for quantization as well as for certain topological effects 
discussed in the last section), we end up with
three contributions. The first one
is precisely the Einstein-Hilbert action (\ref{eh})
written for $R^{ab}$ (which includes
a comological constant, as mentioned), obviously
associated with the gravitational sector. The second one gives
the Yang-Mills action (with the correct relative sign when
considered as a source for gravity) of the
unbroken $SU(2)\times U(1)$ gauge theory.
The third contribution describes 
a set of novel gravity-induced electroweak
processes. We will omit the
details of this calculation because they can be fully
recovered from the case with spontaneous breaking of the
gauge symmetry, to which we now turn.

\section{Spontaneous breaking of the gauge symmetry}

Let us recall a fundamental symmetry of the 
Einstein-Hilbert action (\ref{eh}), by which the latter
remains invariant if the connection 
$\omega_{ab}$ is changed by {\em any}
1-form $\lambda$ to
\begin{equation}
\omega ^{ab} \rightarrow \omega^{ab}+\lambda g^{ab}.
\label{l}
\end{equation}
If $\lambda$ is real, as we will assume, 
this special kind of a projective
transformation is known as Einstein's
$\lambda$ transformation \cite{1}. Obviously, the new connection
violates metricity as
\begin{equation}
Dg_{ab} = -2\lambda g_{ab},
\label{m}
\end{equation}
which apparently is the reason why $\lambda$-transformations 
have rarely been used in recent times (see, however, \cite{5}).
Thus uncovered,
a degeneracy of the vacuum exists in the sense that the same 
Einstein-Hilbert action (\ref{eh}) describes not only
the `symmetric vacuum' $M^4$ corresponding to
$\lambda=0$, but equally well any other vacuum $M_\lambda^4$
with any $\lambda\neq 0$. Remarkably enough,
if we repeat the construction of the previous section
not in $M^4$ but rather in
$M_\lambda^4$, then the gauge symmetry of the Yang-Mills
sector is broken. Conforming to standard
terminology, we may characterize the
breaking as {\em spontaneous}. The magnitude of this breaking
is directly proportional to the scale of $\lambda$ which,
at the outset, has nothing to do with the Planck scale.

To see explicitly how this mechanism works, 
we will repeat the steps leading
to (\ref{R}), but now we must complexify
starting with the connection (\ref{l}) for
the vacuum $M_\lambda^4$. In other words,
instead of $\tilde{\omega}^{ab}$, 
we must employ the connection
\begin{equation}
\tilde{\omega}_{(\lambda)}^{\;\;\;ab}= 
\omega^{ab}+
\lambda g^{ab}+iK_{(\lambda)}^{\;\;\;ab}.
\label{oml}
\end{equation}
In this expression we have
anticipated that the contorsion will change (while
retainning its antisymmetry), as it
indeed does described by
\begin{equation}
K_{(\lambda)\cdot b}^{\;\;\;\;\;a}\wedge e^b =
-\gamma^I_J\theta^a_I\left(F^J
+\lambda\wedge A^J\right).
\label{kl}
\end{equation}
This result, which obviously replaces the previous
expression (\ref{k}), follows from the new set of Cartan's
equations which replace (\ref{tor},\ref{R}) as
\begin{eqnarray}
\tilde{T}^{\;\;\;a}_{(\lambda)}&=&
T^a+\lambda \wedge e^a+\gamma^I_J\theta_I^bA^J\wedge 
K^{\;\;\;\;\;a}_{(\lambda)\cdot b} 
-\gamma^I_J\theta^a_I\lambda \wedge A^J
+i{\cal O}^a
\label{torl} \\
\tilde{R}^{ab}&=& R^{ab}+g^{ab}D\lambda
-K^{\;\;\;\;\;a}_{(\lambda)\cdot c}\wedge  
K^{\;\;\;cb}_{(\lambda)}+ 
\frac{1}{2}C^{ab}_{\;\cdot c}
\tilde{T}^{\;\;\;c}_{(\lambda)}
+\frac{1}{4}\gamma^I_J \gamma^K_L
\theta^a_I \theta^b_K A^J\wedge A^L
+i{\cal O}^{ab}
\label{Rl}
\end{eqnarray}
where we have defined
\begin{eqnarray}
{\cal O}^a & \equiv &
\gamma^I_J\left(D\theta^a_I\wedge A^J+
\theta^a_I F^J +
\theta^a_I\lambda \wedge A^J\right)+
K^{\;\;\;\;\;a}_{(\lambda)\cdot b}\wedge e^b, 
\label{torli} \\
{\cal O}^{ab} & \equiv &
DK^{\;\;\;ab}_{(\lambda)}-
\frac{1}{4}\gamma^I_J(\theta^a_Ie^b+
\theta^b_Ie^a)\wedge A^J. 
\label{Rli}
\end{eqnarray}
The above expressions obviously reduce to (\ref{tor},\ref{R})
at the $\lambda=0$ limit. The most prominent
difference between the two sets is 
the already noted presence of the term proportional
to $\lambda$ in (\ref{kl}). It is precisely this
term which will give masses to the gauge bosons, as we will
see in the next section. The components of
$K_{(\lambda)}^{\;\;\;ab}$
can be calculated explicitly from (\ref{kl}), if we also
take into account the specific value for the $\gamma_I^J$
introduced in (\ref{dot}). The result of this
calculation is
\begin{equation}
2K_{(\lambda)abc}=
\cos \vartheta_W C_{abc}+
\sin \vartheta_W S_{abc},
\label{kle}
\end{equation}
\begin{equation}
C_{abc}=
\theta_{0a}\left(F^0_{bc}+\lambda_bA^0_c-\lambda_cA^0_b\right)+
\theta_{0b}\left(F^0_{ca}+\lambda_cA^0_a-\lambda_aA^0_c\right)-
\theta_{0c}\left(F^0_{ab}+\lambda_aA^0_b-\lambda_bA^0_a\right),
\label{C}
\end{equation}
\begin{equation}
S_{abc}=
\theta_{ja}\left(F^j_{bc}+\lambda_bA^j_c-\lambda_cA^j_b\right)+
\theta_{jb}\left(F^j_{ca}+\lambda_cA^j_a-\lambda_aA^j_c\right)-
\theta_{jc}\left(F^j_{ab}+\lambda_aA^j_b-\lambda_bA^j_a\right),
\label{S}
\end{equation}
wherefrom we may recover, if needed, the explicit values
for the components of
$K^{ab}$ by setting $\lambda =0$.
We thus have at our disposal everything we need to
find explicitly the classical action of the theory.

Again dropping the imaginary surface 
and gauge-fixing terms (essentially the ${\cal O}$
terms in (\ref{torl},\ref{Rl})),
we obtain the Einstein-Hilbert action
\begin{equation}
I_{(\lambda)EH} \equiv \frac{1}{32\pi G}
\int_{M_{\lambda}^4}\tilde R_{(\lambda)ab}\wedge\ast(e^a\wedge e^b)
\label{El}
\end{equation}
expressed as
\begin{equation}
I_{(\lambda)\;EH} = 
\int_{M_\lambda^4}
{\cal L}_{EH}+
{\cal L}_{YM}+
{\cal L}_{gw}+
{\cal L}_{mass}.
\label{ehyml}
\end{equation}
The first contribution in (\ref{ehyml})
describes the gravitational sector with
\begin{equation}
{\cal L}_{EH}=\frac{1}{32\pi G}
R_{ab}\wedge\ast(e^a\wedge e^b),
\label{ehymg}
\end{equation}
reproducing precisely the action (\ref{eh}).
The second contribution gives the Yang-Mills action for 
the $SU(2)\times U(1)$ gauge field with
\begin{equation}
{\cal L}_{YM}=\left(
\frac{\cos^2 \vartheta_W}{2g^2}F^{0ab}F^0_{ab}+
\frac{\sin^2 \vartheta_W}{2g^2}F^{jab}F^j_{ab}\right)
e^0\wedge e^1\wedge e^2\wedge e^3,
\label{ehymk}
\end{equation}
and the indicated couplings.
The third contribution is too long
(to be in good taste and context), so it will
be described only formally as
\begin{equation}
{\cal L}_{gw}=
\frac{1}{g^2}\left(\frac{1}{2L^2}S^{ab}_{a'b'}
\theta_{I}^{a'}\theta_{J}^{b'}A^{I}_{a}A^{J}_{b} +
\frac{1}{4}T^{ab}_{a'b'}
\theta_{I}^{a'}\theta_{J}^{b'}F^{I}_{ac}F^{Jc}_{b} 
\right)+\frac{1}{gL\sqrt{32\pi G}}U^{Kabc}_{\;a'b'c'}
\theta_{I}^{a'}\theta_{J}^{b'}\theta_{K}^{c'}A^{I}_{a}F^{J}_{bc}. 
\label{gw} 
\end{equation}
It predicts a novel kind of electroweak processes 
induced by gravity (we will call them
gravitoweak) which are in fact of two generic types.
The first one 
involves couplings
of the order of $g^2$ and $L^2$ (the Planck scale cancels out
in view of (\ref{G})). The second type, described by the
last term in (\ref{gw}), is at a much higher
scale, roughly given by the geometric mean of $L$ (the
electroweak breaking scale) and the Planck
scale, namely at about $10^{10} Gev$.
In general, the couplings
(and scattering amplitudes) specified by the
matrices $S,T,U$ are fully calculable,
but the explicit tree-level predictions which are thus available
will be meaningful only after the $A_I$'s have been
rotated to physical states. This rotation 
will be effected by $\Delta_I^J$, the
diagonalizer of the mass matrix $M_{IJ}$, 
the latter specifying the last
contribution in (\ref{ehyml}) as
\begin{equation}
{\cal L}_{mass}=
\frac{1}{g^2L^2}M_{IJ}(\lambda,\theta, \vartheta_W)A^{Ia}A^J_a
e^0\wedge e^1\wedge e^2\wedge e^3.
\label{mm} 
\end{equation}
$M_{IJ}$ is fully expressible in 
terms of the parameters $\lambda, \theta_I$ and $\vartheta_W$
($L$ is the scale of $\lambda$), which
fully determine the symmetry-breaking pattern,
as we will see in the next section.
The contribution (\ref{mm}) can be viewed as
a gravitational substitute 
for the Higgs sector. The latter 
is hereby abrogated because we have no option
on the spontaneous breaking of the gauge symmetry
presented in this section: it is imposed by the
geometry and the only existing freedom is in the
choice of the scale and orientation of $\lambda$.

\section{Relevance to the electroweak interaction}

With the generic mixing 
introduced by (\ref{dot}),
the mass matrix as defined by (\ref{mm}) 
turns out to be
\begin{eqnarray}
M_{00}&=&\theta_0^a\theta_0^b
(\lambda^2g_{ab})
-\lambda_a\lambda_b)
\cos^2\vartheta_W, \nonumber\\
M_{0i}&=&M_{i0}=-\theta_0^a\theta_i^b
\lambda_a\lambda_b\sin\vartheta_W\cos\vartheta_W, \nonumber\\
M_{ij}&=&\theta_i^a\theta_j^b
(\lambda^2g_{ab}
-\lambda_a\lambda_b)
\sin^2\vartheta_W.
\label{mmc} 
\end{eqnarray}
More explicit values will obviously
depend on the particular choice of $\lambda$. For
example, we can orient $\lambda$
(without loss of generality) to lie
in the $(\theta_0,\theta_1)$ plane as
\begin{equation}
\lambda^a = \frac{g}{L\sqrt{32\pi G}}(l_0\theta_0^a+
l_1\theta_1^a),
\label{l01} 
\end{equation}
so, with $L$ and
the ratio $l_1/l_0$ (or $l_W/l$) as two
independent real constant parameters, we have
\begin{equation}
\lambda^2 = \lambda^a\lambda_a=
\frac{l_0^2+l_1^2}{L^2} \equiv 
\frac{l^2}{L^2}, \;\;
l_W^2\equiv
l_0^2\sin^2 \vartheta_W+ 
l_1^2\cos^2 \vartheta_W. 
\label{ll} 
\end{equation}
From (\ref{mmc},\ref{l01}), we find
for the mass matrix and its diagonalizer
\begin{equation}
M_{IJ}=
\left(\begin{array}{cccc}
l_1^2\cos^2 \vartheta_W &
-l_0l_1\sin \vartheta_W \cos \vartheta_W &0&0\\
-l_0l_1\sin \vartheta_W \cos \vartheta_W 
&l_0^2\sin^2 \vartheta_W &0&0\\
0&0&l^2\sin^2 \vartheta_W&0 \\
0&0&0&l^2\sin^2 \vartheta_W
\end{array}\right),
\label{mmw}
\end{equation}
\begin{equation}
\Delta_{I}^{J}=
\left(\begin{array}{cccc}
(l_0/l_W)\sin \vartheta_W &
(l_1/l_W)\cos \vartheta_W &0&0\\
-(l_1/l_W)\cos \vartheta_W &
(l_0/l_W)\sin \vartheta_W &0&0\\
0&0&1&0\\
0&0&0&1
\end{array}\right).
\label{delta}
\end{equation}
Taking also into account the scale from (\ref{mm}), we obtain the
mass spectrum
\begin{equation}
m_0^2=0,\;\;
m_1^2 = \frac{l_W^2}{g^2L^2}
\equiv m_Z^2,\;\; 
m_2^2 = m_3^2 = \frac{l^2}{g^2L^2}
\sin^2 \vartheta_W
\equiv m_W^2,
\label{dmm}
\end{equation}
obviously acquired by the physical bosons
\begin{equation}
B=\frac{l_0}{l_W}\sin \vartheta_WA^0+ 
\frac{l_1}{l_W}\cos \vartheta_W A^1,\;\;
Z=-\frac{l_1}{l_W}\cos \vartheta_W A^0+
\frac{l_0}{l_W}\sin \vartheta_W A^1,\;\;
W^{+}=A^2,\;\;W^{-}=A^3.
\label{pb}
\end{equation}
We can trade $L$ 
for $m_W$ (or $m_Z$) 
and $l_W/l$ 
(or $l_1/l_0$) 
for the positive parameter
\begin{equation}
\rho \equiv \frac{m_Z^2\cos^2 \vartheta_W}{m_W^2}=
\frac{l_0^2\sin^2 \vartheta_W+l_1^2
\cos^2\vartheta_W}{(l_0^2+l_1^2)\tan^2 \vartheta_W}=
\left(\frac{l_W}{l\tan \vartheta_W}\right)^2,
\label{rho}
\end{equation}
so that, following the $\rho=1$ {\em choice} (see,
however, following remarks), we
recover precisely the mass spectrum 
of the standard electroweak model.
One may now proceed to fully determine the gravitoweak
sector given by (\ref{gw}), 
which however is beyond our present scope and will be 
examined elsewhere.

It may have been already noticed that our construction 
carries certain aspects of a Kaluza-Klein setting \cite{2g}.
These could be profitably exploited in spite of obvious
fundamental differences in
dimensionality or the fact that ${\cal P}$ 
is not necessarily an isometry group 
(although it is expected to be so 
asymptotically, e.g., in models which attain the homogenization
mentioned earlier). In any case, 
there seems to be no obstruction in
applying existing methods 
to obtain results such as the quantization
of the electric charge and the computability of 
$g$,$\vartheta_W$,$\rho$ \cite{w}. As related
to that, we may already demonstrate in the present context an elegant
formulation of the generalized minimal coupling
prescription, which
is precisely carried by $\tilde{e}_a$ (the dual of
$\tilde{e}^a$) and automatically assigns the correct
charges to the electroweak gauge bosons.
For an explicit expression let us assume that the
imaginary part of $\tilde{e}^a$ in (\ref{et}) is small
so that, to lowest order in $h$, we have
\begin{equation}
\tilde{e}_a=(\delta^b_a+
i\gamma^I_J\theta^b_I A^J_a)e_b,
\label{pa}
\end{equation} 
where we should actually rotate the $A_I$ to the
physical states (\ref{pb}). 
To better recognize this result, one may convert
to holonomic co-ordinates and disregard the
non-abelian contributions. Then, (\ref{pa}) further reduces 
to the electromagnetic minimal coupling prescription
$\partial_a-ie A_a$, with the electric charge $e$ emerging
through the identification of the relevant charge operator.

\section{Discussion and conclusions}

Shortly after its discovery as a possible
space-time symmetry, the de Sitter group has been 
repeatedly advocated as an option superior
compared to the Poincar\'{e} group \cite{ds}.
Here we have seen that the
complexified gauging of the de Sitter
group, albeit with its algebra expressed in the 
particular basis (\ref{PP}) for the isomorphic ${\cal P}$,
has uncovered a unified description of the gravitational
and electroweak interactions 
(cf. also comments at the begining of section 2).
The construction is fixed
by less than five
adjustable parameters (at best only the two scales) 
among the $G,g,m_W,
\vartheta_W,\rho$. We have seen that
the known association of gravity with the 
generators of the Lorentz group ${\cal L}$ 
has been retained, while
the $\Pi_a$
have been uniquely associated with
the electroweak interaction. Our findings
offer new theoretical perspectives and
predictions which apparently could upgrade general relativity
as well as the standard electroweak
model. The implied
programme is obviously vast so we will only list
what appear at the moment as its major aspects, 
also to be thought of as
testing grounds on its worthiness.

To the extend that one can isolate gravity, Einstein's
theory remains unchanged exept for one major issue, namely
the dramatic reduction 
of the immense body
of all possible {\em global}
topologies \cite{6} to just
${\cal P}/{\cal L}$.
Einstein's equations (possibly 
with the addition of appropriate external sources) will still
have to be solved, and even the asymptotic-flatness boundary
conditions may certainly retain most (but not all) of their
practical applicability, e.g., for an isolated local 
source. Clearly however, solutions
with topology consistent to that of
${\cal P}/{\cal L}$ (which includes several models with
Bianchi-type 
symmetry \cite{6}) would be of special
interest within the present context. Transcending the gravitational 
sector, novel
effects are expected such as those related to
violations of metricity and to alterations
of the general relativistic
junction conditions on the surface
of appropriately chosen sources. The former should be observable
through tests for violation of the principle of equivalence,
expected to be positive if their accuracy
exceeds one part
in $10^{17}$
(the ratio of the electroweak
to the Planck scale).
The second type uncovers a new 
generation of Aharonov-Bohm -like effects 
and a novel insight for the Blackett effect, 
as already discussed in a related context \cite{8}.

Our findings also provide a unification of
the electroweak sector when considered by itself.
The association of the group generators with a 
space-time vierbein (which could offer an
elegant explanation of the origin and
uniqueness of the $SU(2)\times U(1)$ choice)
is in no way contradicted by any 
rigorous result in the Coleman-Mandula theorem, 
but it does supply a counter-example to some
fundamental assumptions therein \cite{cm}. 
The gravitoweak sector
involves, as we have seen, two types of 
gravity-induced electroweak interactions with
couplings at a low (essentially electroweak)  scale and at
an intermediate scale of about
$10^{10}Gev$.

On major open issues, we note that the 
modifications in the commutation relations (\ref{ccp}),
e.g., the association of
the $[\Pi_j,\Pi_0]$ with a Lorentz boost
along $j$, will be clarified
once the representations of ${\cal P}$ (with
the related kinematics etc.) have been worked out
in full detail. We also note
that the predicted absence of a Higgs sector,
congruous as it may be with current doubts,
will expose the electroweak sector to contamination by
the renormalization and quantization impasses
for gravity, possibly made worse by unitarity
problems. Contributions to the latter could also 
come from the Goldstone-boson analogue expected
from the symmetry breaking: although
Einstein's $\lambda$-transformation is an
invariance of the classical vacuum, it does 
contribute with surface and gauge terms, as seen
from (\ref{Rl}). 

Such contributions can also be expected from terms
in (\ref{torl}-\ref{Rli}) (e.g. the ${\cal O}$'s)
which were dropped
in the classical action (\ref{ehyml}). On the other
hand, the same terms seem to convey a welth of
topological configurations and novel effects,
including the classical ones mentioned above.
Additional (and possibly exploitable) novel aspects
may be itemized as follows:
the simple group ${\cal P}$ replacing 
the (essentially responsible for the mentioned
impasses) ${\cal P}_0$; novel insight on 
the CPT theorem and chiral behavior, or on
spin and space-time parallelizability,
through their association (together with the
electroweak bosons)
with the globally defined
$\theta^a$-frame
\cite{cp};
the uncovered gravitoweak processes and novel effects;
the rigorous foundation of the minimal coupling 
prescription offered by $\tilde e_a$;
the expected predictability and protection of the
$g,\vartheta_W$, 
$\rho=1$ values; and the hereby anticipated explanation of
Dirac's large-number conjecture. We finally note
the rather unexpected (and certainly peculiar) synthesis of 
several major theoretical aspects:
grand-unification with ${\cal P}$ (albeit
geometrical); 
superseding of the Coleman-Mandula theorem (without
supersymmetry);
spontaneous breaking of the gauge symmetry (without Higgs 
fields); complexification
(without complex or twistor structures \cite{1});
and Kaluza-Klein aspects (without
extra dimensions!).
Amusing as it may be, this
assortment may also supply options to be further pursued.

\end{document}